# Exceptional Cones in 4D Parameter Space

Qiang Wang[1,4*], Kun Ding[2, 3*], Hui Liu[1†], Shining Zhu[1], and C. T. Chan[2*]

[1]*National Laboratory of Solid State Microstructures, School of Physics, Collaborative Innovation Center of Advanced Microstructures, Nanjing University, Nanjing 210093, China*

[2]*Department of Physics and William Mong Institute of Nano Science and Technology, The Hong Kong University of Science and Technology, Clear Water Bay, Kowloon, Hong Kong*

[3] *The Blackett Laboratory, Department of Physics, Imperial College London, London SW7 2AZ, United Kingdom*

[4]*Division of Physics and Applied Physics, School of Physical and Mathematical Sciences, Nanyang Technological University, Singapore 637371, Singapore*

* These authors contributed equally.

† Corresponding author: liuhui@nju.edu.cn

* Corresponding author: phchan@ust.hk

**Abstract**

The notion of synthetic dimensions has expanded the realm of topological physics to four dimensional (4D) space lately. In this work, non-Hermiticity is used as a synthetic parameter in PT-symmetric photonic crystals to study the topological physics in 4D non-Hermitian synthetic parameter space. We realize a 3D exceptional hypersurface (EHS) in such 4D parameter space, and the degeneracy points emerge due to the symmetry of synthetic parameters. We further demonstrate the existence of exceptional degenerate points (EDPs) on the EHS that originates from the chirality of exceptional points (EPs), and the exceptional surface near EDPs behaves like a Dirac cone. We further show that a very narrow reflection plateau can be found near these EDPs, and their sensitivity towards the PT-symmetry breaking environmental perturbation can make these degeneracy points useful in optical sensing and many other nonlinear and quantum optical applications.

## 1. Introduction

Topological physics in synthetic dimensional space[1] is attracting more and more attention lately. Although the real physical space only has three spatial dimensions, higher-dimensional topological properties can in principle be defined and explored using synthetic dimensions. Recent examples include four-dimensional quantum hall effect[2-4], synthetic gauge fields[5], 5D Weyl semimetal[6], Wilson Loops[7], high order topological insulator [8, 9], 3D topological insulators [10] and so on[11, 12]. In addition, 3D Weyl point [13] and topological insulators in 2D microcavity array[14] have been realized in synthetic space. On the other hand, one-dimensional photonic crystals provide a very simple and practical experimental platform to investigate topological properties of energy bands up to optical frequencies. Recently, through defining geometry parameters as synthetic dimensions, Weyl points were obtained in 3D parameter space[15, 16] and interesting phenomena such as reflection phase vortex and the related topological interface states ("fermi arc") were observed in optical frequencies.

In the meantime, non-Hermitian systems and in particular, PT-symmetric systems, are attracting more recent attention [17]. In optical structures, PT-symmetric photonic modes can be obtained through adding gain and loss in materials, and have been used to realize unidirectional waveguides[18, 19] and lasing[20-22]. The PT-symmetric phase transition can also induce coalescence of both eigenvalues and eigenvectors, leading to degeneracies called exceptional points (EPs)[17, 23, 24]. The novel properties at and near EPs can be used for various applications such as optical sensing[23-25], nonlinear optics[26, 27] and slow light devices[28, 29]. Adding gain and loss to topological materials can create a new platform to observe new phenomena, in particular those associated with degeneracies and singularities [30-35]. In general, the degeneracy points of Hermitian topological bands can be changed to EP structures of non-Hermitian topological bands. For examples, diabolic points can be changed to two EPs[36], Weyl points can be changed to EP rings [37-39] and a nodal line can be changed to an EP surface (ES) [40, 41].

In this work, we propose a simple platform that is rich in physics and can be used to investigate PT-symmetric topological physics in 4D. Here, employing a 1D photonic crystal with a seven-layer unit cell, we construct a 4D parameter space $(p,q,k,\gamma)$ based on wavevector $k$, two geometric parameters $p$, $q$ and one non-Hermitian coefficient $\gamma$. A 3D EHS is obtained in such a 4D space. Through varying the non-Hermitian coefficient, we found two kinds of exceptional degenerate points (EDPs) in the EHS emerge in the process. As photonic gap is closed at EDPs, and a very narrow gap can be introduced near the EDP before the gap's closing. As a result, a narrow reflection peak with high Q factor can be produced near the EDP. A small change of environment which breaks the PT-symmetry will change reflection drastically. This interesting property of EDPs can potentially be used in optical sensing.

## 2. Structure and the nodal line

In previous work, when adding the non-Hermitian perturbation, a Weyl point will change into an exceptional ring[38]. Therefore, introducing non-Hermiticity to the nodal line could deform it to an exceptional surface or not is a worth pursuing topic, and this is not hard to study by introducing the parameter space. In this paper, we first realize the Nodal line in the parameter space. Then we introduce the non-Hermiticity but preserve the PT symmetry of the system, we can clearly see the nodal line will change into exceptional tours.

The unit cell consists of a stack of seven layers incorporating gain and loss, as shown in Fig. 1(a). The thickness of the layers is:

$$\begin{aligned} d_{a1} &= (1+\cos(p\pi))d_a \\ d_{b1} &= (1+\cos(q\pi))d_b/2 \\ d_{a2} &= (1-\cos(p\pi))d_a/2 \\ d_{b2} &= (1-\cos(q\pi))d_b/2 \end{aligned} \quad (1)$$

To be experimentally doable, we set $d_a = 0.3250\ \mu\text{m}$ and $d_b = 0.2414\ \mu\text{m}$, and the

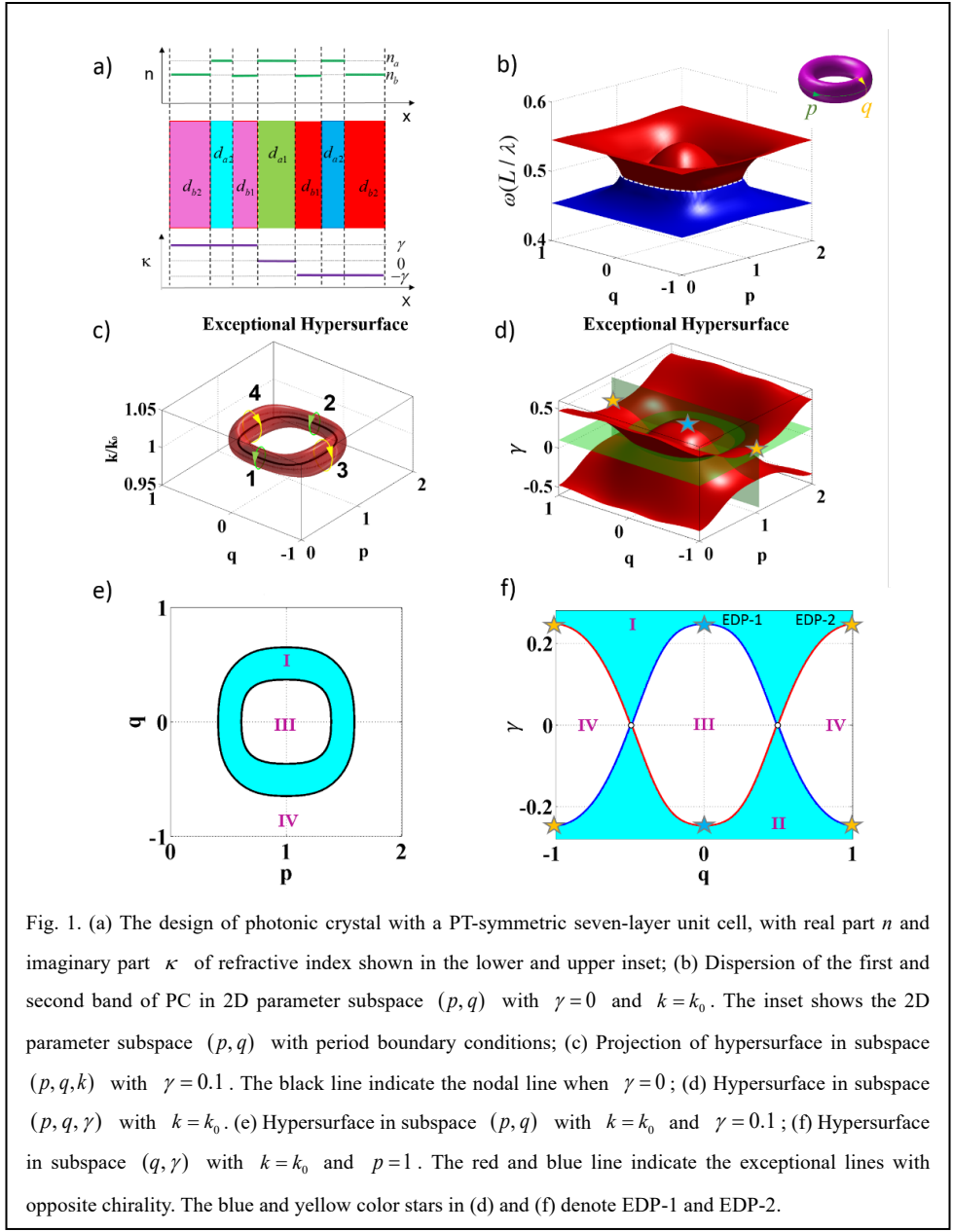

Fig. 1. (a) The design of photonic crystal with a PT-symmetric seven-layer unit cell, with real part $n$ and imaginary part $\kappa$ of refractive index shown in the lower and upper inset; (b) Dispersion of the first and second band of PC in 2D parameter subspace $(p, q)$ with $\gamma = 0$ and $k = k_0$. The inset shows the 2D parameter subspace $(p, q)$ with period boundary conditions; (c) Projection of hypersurface in subspace $(p, q, k)$ with $\gamma = 0.1$. The black line indicate the nodal line when $\gamma = 0$; (d) Hypersurface in subspace $(p, q, \gamma)$ with $k = k_0$. (e) Hypersurface in subspace $(p, q)$ with $k = k_0$ and $\gamma = 0.1$; (f) Hypersurface in subspace $(q, \gamma)$ with $k = k_0$ and $p = 1$. The red and blue line indicate the exceptional lines with opposite chirality. The blue and yellow color stars in (d) and (f) denote EDP-1 and EDP-2.

period of the unit cell is $\Lambda = 2(d_a + d_b)$. Real parts of refractive index $n$ of the seven layers are set as $n_a = 2$ and $n_b = 1.45$ which are mirror-symmetric, while the imaginary parts of refractive index $\kappa$ are anti-mirror-symmetric (Fig. 1(a)). Such a PC possesses PT-symmetry. The non-Hermitian parameter $\gamma$ represent the optical gain and loss inside the material, which is a tunable parameter in this work. To acquire gain material, we can use the semiconductors, such as InGaAsP multiple quantum wells[22]; another way is to dope the rare earth element into the conventional material[42]. For the loss term, we can also dope metallic element to control. The real part of optical length is $L = 2(n_a d_a + n_b d_b)$, which does not depend on the values of $p$ and $q$, so the central frequency of the band gap is almost fixed. The synthetic parameters $(p, q)$, together with the Bloch wavevector $k$ of the PC, constitute a 3D parameter space $(p, q, k)$. The parameter subspace $(p,q)$ possesses mirror symmetry, as $(p,q)$ and $(-p,-q)$ are by definition exactly the same system.

Based on transfer matrix method (TMM), we can calculate the band dispersion $\omega_n(p,q,k)$ of PC in the 3D parameter space, where $n$ denotes the $n$-th band of the PC. For the 1D PC, the first band gap between the first and the second band is located at Brillouin zone boundary $k_0 = \pi/\Lambda$. Therefore, we show the dispersion of these two bands $\omega_n(p,q,k)$ in the 2D $(p,q,k_0)$ subspace, as shown in Fig. 1(b). We see that these two bands indeed have a set of degeneracy points (white dashed line) which form a nodal line, as shown by the black line in Fig. 1(c). The effective Hamiltonian for this nodal line can be retrieved based on the data calculated from transfer matrix as

$$H_0 = d_x \sigma_x + d_y \sigma_y, \qquad (2)$$

with $d_x = 0.020\left(\cos(p\pi)^2 + \cos(q\pi)^2 + 2.4(\cos(p\pi) - \cos(q\pi)) + 1.4\right)$, and $d_y = 0.946(k/k_0 - 1)$. In principle, the Hamiltonian can be derived from TMM[15], but we can also simplify the procedure based on the symmetry of such system near the nodal line. Note that the Hamiltonian has square terms in $p$ and $q$ due to the inversion symmetry of the synthetic space. While in the $k$ dimension, we only keep the linear term, as the crossing of 1D PCs near the nodal line is always linear as shown in Fig.1(b). In Eq. (2), we neglect the $\sigma_0$ term, which only determines the frequency of the nodal line. The nodal line is obtained through $d_x = d_y = 0$. $d_y = 0$ tells the nodal line lies on the $k = k_0$ plane, and $d_x = 0$ shows the nodal line should satisfy $(\cos(p\pi)+1.2)^2 + (\cos(q\pi)-1.2)^2 = 1.48$. It is worthy to mention that the result is quite different from that in the former four-layer PC[15], in which only one single degeneracy point is found between the first and the second band, which is obtained by breaking the inversion symmetry.

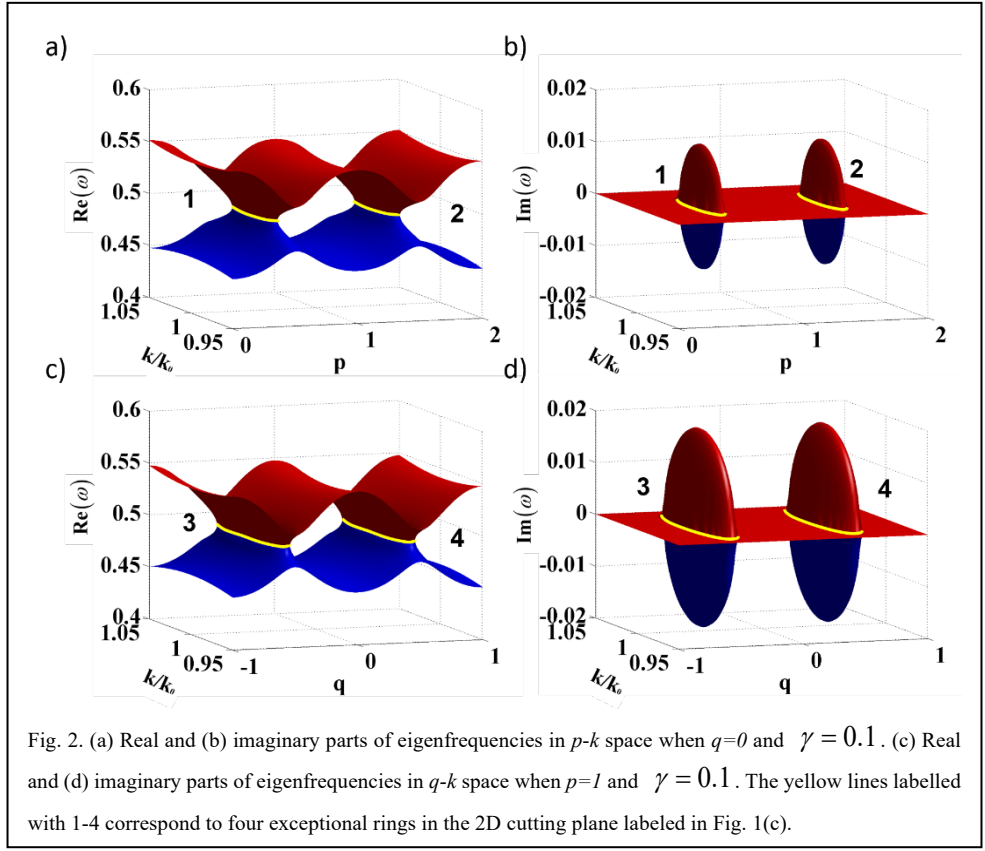

Fig. 2. (a) Real and (b) imaginary parts of eigenfrequencies in $p$-$k$ space when $q=0$ and $\gamma = 0.1$. (c) Real and (d) imaginary parts of eigenfrequencies in $q$-$k$ space when $p=1$ and $\gamma = 0.1$. The yellow lines labelled with 1-4 correspond to four exceptional rings in the 2D cutting plane labeled in Fig. 1(c).

## 3. Exceptional hypersurface and exceptional degenerate point

We already show that a nodal line can exist in the parameter space, and now we study in which manner the non-Hermitian parameter will deform the nodal line. As reported recently [40, 43], after introducing gain and loss, the nodal line likely changes to ES under the protection of symmetries. Therefore, we also use TMM to calculate the EPs in the parameter space $(p, q, k)$ for different $\gamma$. For instance, when $\gamma = 0.1$, the numerical result is given in Fig. 1(c). We see that each point in the nodal line evolves into an exceptional ring encircling this point, and hence all the EPs form a torus surrounding the nodal line [37-40, 44]. The torus separates the space into two parts, the volume inside the torus is the PT-broken phase, which means two bands share the same imaginary parts of eigenvalues; while the volume outside torus is the PT-symmetric phase, and two bands share the same real parts of eigenvalue. To show such dispersions in the ES of Fig. 1(c), we plot in Fig. 2 the real and imaginary parts of eigenfrequencies of 1D PCs in p-k (a, b) and q-k (c, d) space, we can clearly see two exceptional rings highlighted by yellow lines, which is labelled with 1-4 as in Fig. 1(c). On the four exceptional rings, the real and imaginary part of the eigenfrequencies are both degenerate.

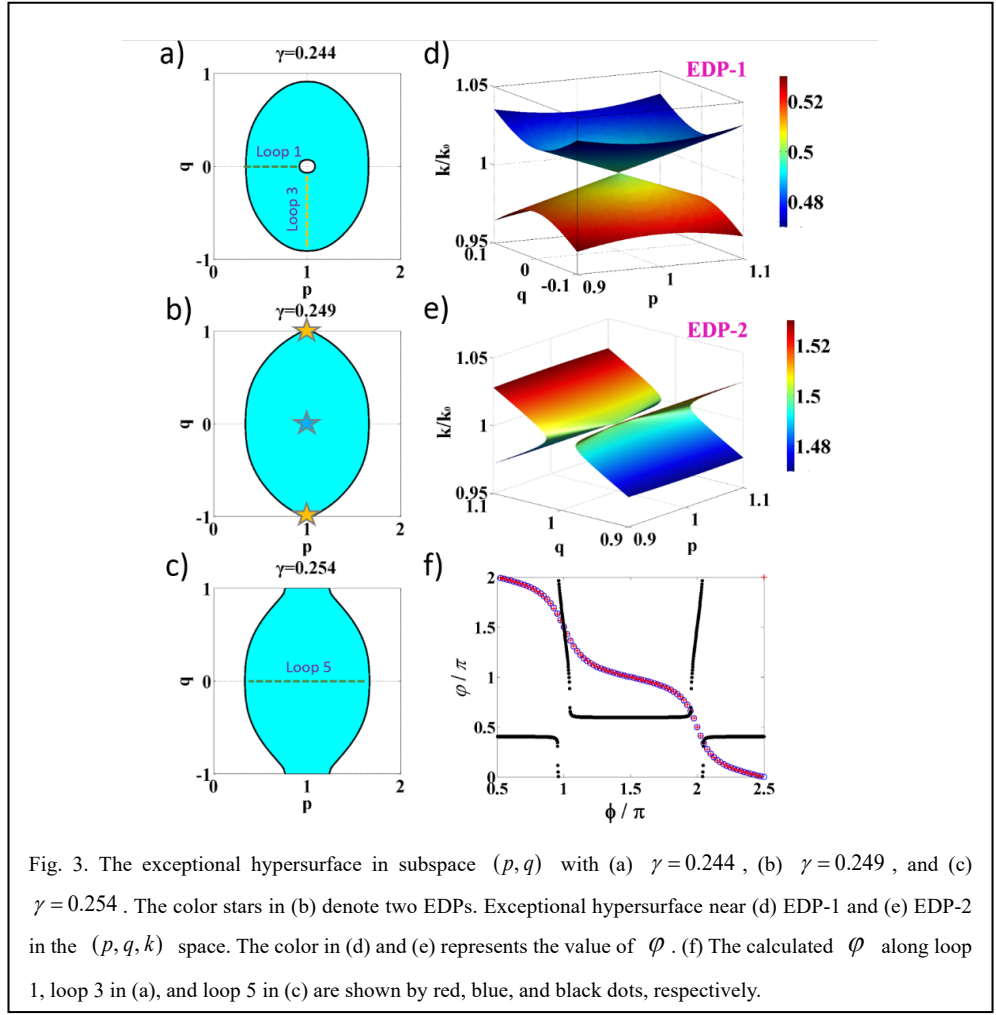

Fig. 3. The exceptional hypersurface in subspace $(p,q)$ with (a) $\gamma = 0.244$, (b) $\gamma = 0.249$, and (c) $\gamma = 0.254$. The color stars in (b) denote two EDPs. Exceptional hypersurface near (d) EDP-1 and (e) EDP-2 in the $(p,q,k)$ space. The color in (d) and (e) represents the value of $\varphi$. (f) The calculated $\varphi$ along loop 1, loop 3 in (a), and loop 5 in (c) are shown by red, blue, and black dots, respectively.

If we treat $\gamma$ as another parameter, then the ES becomes a 3D EHS in the 4D parameter space $(p,q,k,\gamma)$. Figure 1(d) shows such EHS in the parameter space $(p,q,\gamma)$ with $k=k_0$. There are two EHSs intersecting with each other and dividing the 4D space into four regions denoted with I, II, III and IV. The intersection line is the nodal line shown in Fig. 1(c), which lies on the $\gamma=0$ and $k=k_0$ plane. In Figs. 1(e) and 1(f), we plot the EHS on ($\gamma=0.1$ and $k=k_0$) plane and ($k=k_0$ and $p=1$) plane respectively. The two cyan regions I and II are PT-broken phase corresponding to the volume above or below two EHSs in Fig. 1(d), while the other two white regions III and IV are PT-symmetric phase corresponding to two volumes sandwiched between two EHSs in Fig. 1(d). The red and blue lines in Fig. 1(f) represent these two exceptional lines have opposite chiralities.

We already show in Fig. 1(c) the EHS looks like a donut when $\gamma=0.1$, and the cross section in the $(p,q,\gamma=0.1)$ plane is a double ring, as shown by black lines in Fig. 1(e). The increase of $\gamma$ will make the region III and IV gradually disappear, as shown in Fig. 1(f). To see this, we plot the EHS in the 2D $(p,q)$ subspace at three different values of $\gamma$ in Figs. 3(a-c). When $\gamma=0.244$, the EHS is a donut as shown in Fig. 1(c). Increasing $\gamma$ will expand the donut, and the inner ring of EHS will approach the center $(p=1,q=0)$. When $\gamma=0.249$, the inner ring will collapse at the center, shown by blue stars in Fig. 1(d) and 3(b), which we refer it as EDP-1: $(p_1,q_1,k_1,\gamma_1)=(1,0,1,0.249)$. As $\gamma>0.249$, the inner ring disappears, as shown in Fig. 3(c) at $\gamma=0.254$. During the process, the topology of EHS changes, which will be discussed later. The outer ring of EHS approaches to another point $(p=1,q=\pm1)$ at the same value of $\gamma$, as shown by the yellow star in Figs. 1(d) and 1(f), and we refer it as EDP-2: $(p_2,q_2,k_2,\gamma_2)=(1,\pm1,1,0.249)$. Both EDPs locate at the high symmetry points in the parameter space and share the same values of $k$ and $\gamma$ ($k_1=k_2$ and $\gamma_1=\gamma_2$) because these two PCs at the transition points are essentially the same with mirror symmetric unit cells. However, the PCs near the two EDPs have different behavior, which make the exceptional cones (EDP-1 and EDP-2) quite different. To make it clear, we plot the dispersion of PCs near two EDPs in Fig. 4 when $\gamma=0.249$. Figures 4(a) and 4(b) show real and imaginary parts of eigenfrequencies near EDP-1. The black, red and blue line (markers) indicate the PCs with (p, q) equal to (1, 0), (1, 0.1) and (1.1, 0), and the positions are marked with blue pentagon, blue circle and red circles in the insert of Fig. 4(a). Figures 4(c) and 4(d) show real and imaginary parts of eigenfrequencies near EDP-2, and black, red and blue lines (markers) indicate the dispersion of PCs with (p, q) equal to (1, 1), (1, 0.9) and (1.1, 1) marked with yellow pentagon, blue circle and red circles. Comparing these two cases, we can see the dispersion of the PCs at EDP-1 (1, 0) and EDP-2 (1, 1) are the same, which is guaranteed by the symmetry of the parameter space. However, the PCs near these two EDPs show different behavior. We see that there are EPs along the *p* and *q* axis near

EDP-1, while for EDP-2, we can find the EPs only along $q$ axis, but produce a gap along $p$ axis.

## 4. Hamiltonian and topological properties

To study topological properties of EDPs, we discuss Hamiltonian near these EDPs and then investigate geometric phase associated with the point. Figure 3(d) shows the shape of EHS in the $(p,q,k)$ subspace near EDP-1 for $\gamma = 0.249$. We see that the two EHSs linearly touch at one point, and form an exceptional cone. To describe its behavior, we expand the parameters near EDP-1 as $\tilde{p}=p-p_1$, $\tilde{q}=q-q_1$, $\tilde{k}=k-k_1$, and $\tilde{\gamma}=\gamma-\gamma_1$, and then effective Hamiltonian is

$$\mathrm{H}\left(\tilde{p},\tilde{q},\tilde{k},\tilde{\gamma}\right)=d_x\sigma_x+d_y\sigma_y+id_z\sigma_z, \tag{3}$$

where $d_x = 0.0548\tilde{p}^2 + 0.1149\tilde{q}^2 - 0.3758$, $d_y = 1.0013\tilde{k}$, and $d_z = 0.3758 + 0.0940\tilde{\gamma}$. Note here, we keep the square term of p and q due to the inversion symmetry of parameter space. While in 1D PCs, the dispersion near the degenerate point is linear, so here we keep the first order of k. Usually, we can get the Hamiltonian from TMM[15], however we can also simplify the procedure by the symmetry of the system. Equation (3) shows the dispersion near EDP-1 has the following form

$$\omega=\pm\sqrt{-c_1\tilde{p}^2-c_2\tilde{q}^2+c_3\tilde{k}^2-c_4\tilde{\gamma}} \quad, \tag{4}$$

where $c_1 = 0.0412$, $c_2 = 0.0864$, $c_3 = 1.0027$, and $c_4 = 0.0706$. When $\tilde{\gamma} = 0$, the change of EHS at the EDP-1 can be seen from the right side of Eq. (4), and we can directly get $\tilde{k} = \pm \frac{1}{c_3}\sqrt{c_1 \tilde{p}^2 + c_2 \tilde{q}^2}$. It can be seen that the EHS near EDP-1 is a linear cone.

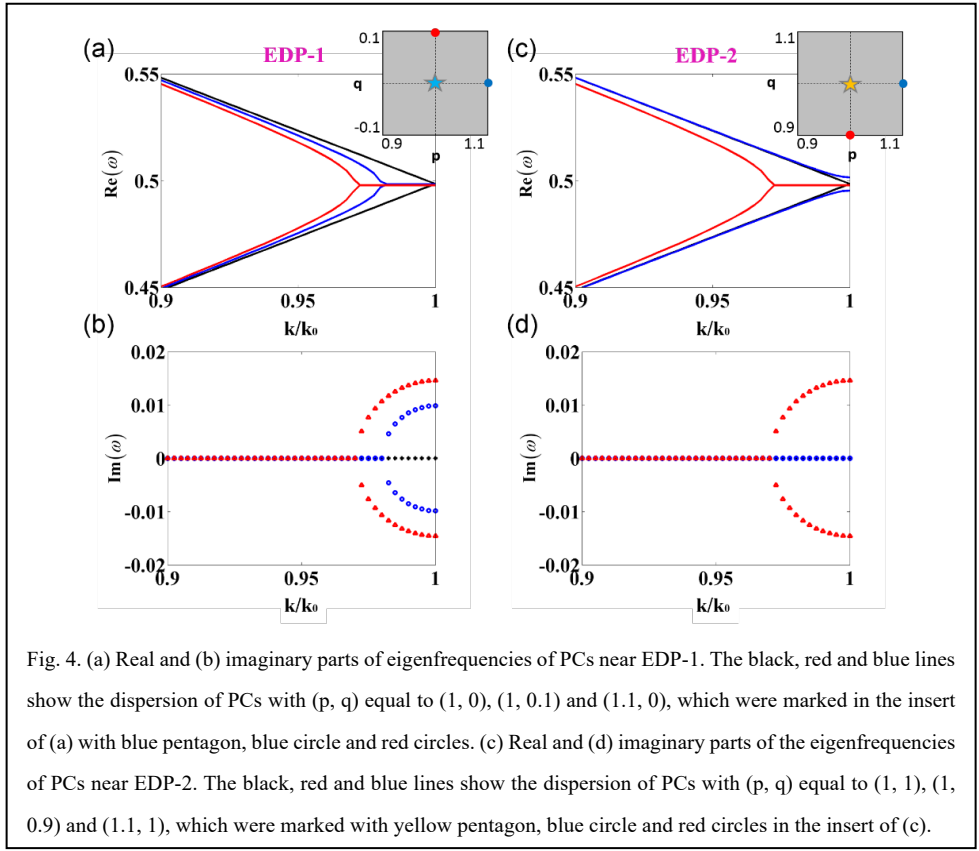

Fig. 4. (a) Real and (b) imaginary parts of eigenfrequencies of PCs near EDP-1. The black, red and blue lines show the dispersion of PCs with (p, q) equal to (1, 0), (1, 0.1) and (1.1, 0), which were marked in the insert of (a) with blue pentagon, blue circle and red circles. (c) Real and (d) imaginary parts of the eigenfrequencies of PCs near EDP-2. The black, red and blue lines show the dispersion of PCs with (p, q) equal to (1, 1), (1, 0.9) and (1.1, 1), which were marked with yellow pentagon, blue circle and red circles in the insert of (c).

For EDP-2 shown in Fig. 3(e), the effective Hamiltonian will have the same form as Eq. (3) with different parameters as $d_x = 0.0067 \tilde{p}^2 - 0.1463 \tilde{q}^2 + 0.2926$, $d_y = -1.0013 \tilde{k}$ and $d_z = 0.2926 + 0.1208 \tilde{\gamma}$. Straightforward, the dispersion near EDP-2 is also given by Eq. (4), but the coefficients are $c_1 = -0.0039$, $c_2 = 0.0856$, $c_3 = 1.0027$, and $c_4 = 0.0706$. Comparing these with those in EDP-1, we see that the values of $c_2$, $c_3$, and $c_4$ are almost the same, but the sign of $c_1$ changes. This sign flip makes the direction of the cone switch to horizontal axis ($\tilde{q} = \pm \frac{1}{c_2}\sqrt{c_1 \tilde{p}^2 + c_3 \tilde{k}^2}$) although the conic shapes of EHS near EDP-1 and EDP-2 are similar.

To see the difference between EDPs in Fig. 3 with the linear crossing in Fig. 1(f), we need to analysis the eigenvector of EP in the EHS [45, 46]. The eigenvector at the EP can be formally written as

$$|u\rangle = \begin{pmatrix} \cos(\theta) \\ \sin(\theta)e^{i\varphi} \end{pmatrix}. \quad (5)$$

Here $\theta$ and $\varphi$ are two independent variables. Since every point in EHS satisfies $d_x^2 + d_y^2 = d_z^2$, then $\theta = \pi/4$, and $\varphi = -\text{sgn}(\gamma)\pi/2 + \text{Arg}(d_x + i d_y)$. In Figs. 3(d) and 3(e), we plot the values of $\varphi$ in the EHS by colors. We can see the $\varphi$ changes smoothly on the surface, and the eigenvector of two EDPs are $|u\rangle = (1, i)/\sqrt{2}$ and $|u\rangle = (1, -i)/\sqrt{2}$, respectively. Note that the linear crossing point here are still EPs due to $d_z \neq 0$ at these points. This is the consequence of coalescence of EPs with the same chirality [45]. The EDPs, which labelled by blue and red stars, have opposite chirality. However, the linear crossing at $\gamma = 0$ in Fig. 1(f) is not an EP becasue the EPs with opposite chirality (blue and red lines in Fig. 1(f)) coalesce at the nodal line.

The topological properties of EDPs can be also confirmed through calculation the change of geometric phase before and after the transition point. When $\gamma = 0.244$, we choose the yellow loop 1 and green loop 3 as shown in Figs. 1(c) and 3(a), and the values of $\varphi$ along these two loops are shown by blue circles and red crosses in Fig. 3(f), respectively. The $\phi$ represents the polar angle according to the center of the two loops, and the arrows in Fig. 1(c) show the directions. We can see that the change of geometric phase on loop 1 and 3 is $2\pi$; while the change of geometric phase on loop 2 and 4 is $-2\pi$. When the exceptional tours expand, the loop 1 and 2 will touch at $(p=1, q=0)$, while the loop 3 and 4 will touch at $(p=1, q=\pm1)$, as shown in Fig. 3(b). When $\gamma = 0.254$, loop 1 and loop 2 will compromise as a big loop 5, as shown in (c), and the change of geometric phase on this loop is 0 as shown in Fig. 3(f) by black dots. This confirms the EDP is the topological transition point in the EHS, across which the genus of the exceptional surface changes.

## 5. Reflection of PCs with finite periods

It is known that Dirac cone in band structure can lead to many interesting properties, such as zero-index medium[47], Klein tunneling[48], etc. For exceptional points, there are also some potential of application have be proposed, such as realizing unidirectional waveguides[18, 19], and lasing[20-22] . Usually, the reflection or transmission spectra is widely used in sensors[49, 50], in this section, we will show the EDPs can also be used as sensors by measuring the refection spectrum, which is quite sensitive to the change of refractive index in the surrounding.

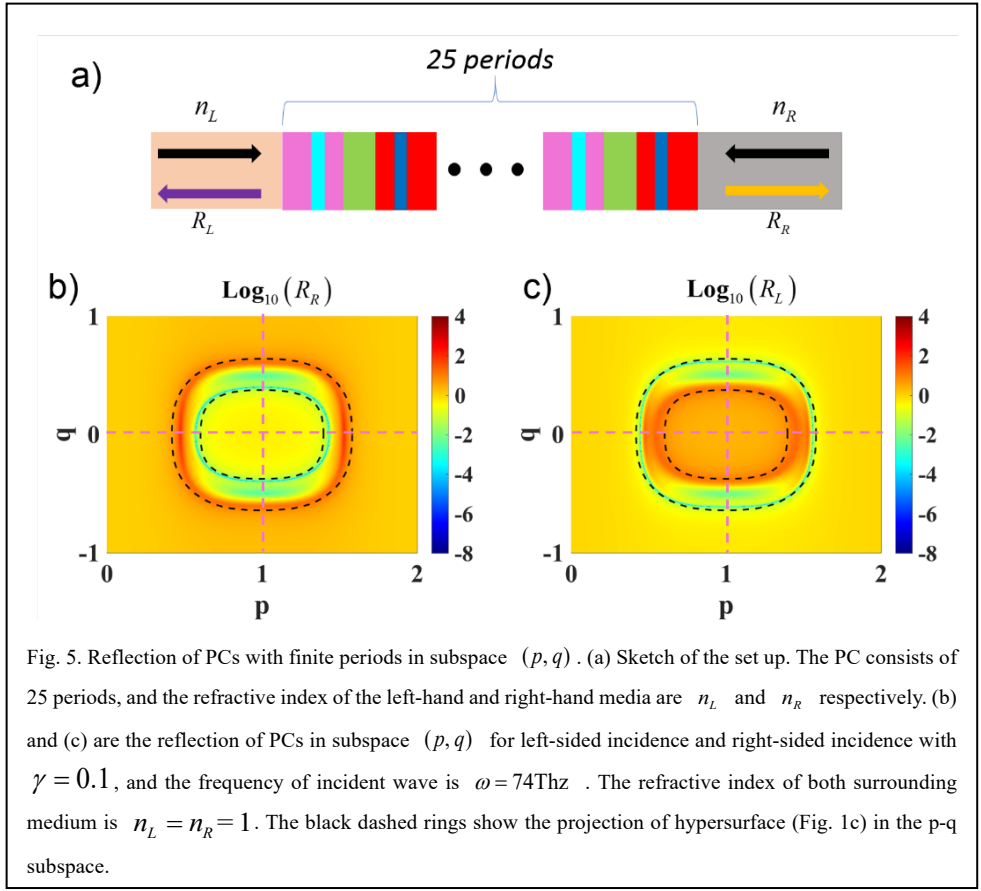

Fig. 5. Reflection of PCs with finite periods in subspace $(p,q)$. (a) Sketch of the set up. The PC consists of 25 periods, and the refractive index of the left-hand and right-hand media are $n_L$ and $n_R$ respectively. (b) and (c) are the reflection of PCs in subspace $(p,q)$ for left-sided incidence and right-sided incidence with $\gamma = 0.1$, and the frequency of incident wave is $\omega = 74\text{Thz}$. The refractive index of both surrounding medium is $n_L = n_R = 1$. The black dashed rings show the projection of hypersurface (Fig. 1c) in the p-q subspace.

In the following, we first investigate the transmission and reflection properties of the PCs to confirm the existence of EDPs. We design a finite PC with 25 unit cells shown in Fig. 5(a). The refractive index of environment medium on the left and right side are $n_L$ and $n_R$ respectively. The whole system is PT-symmetric when $n_L = n_R$ and PT-symmetry broken when $n_L \neq n_R$. We first study transport properties near the EHS when $n_L = n_R$. Figures 5(b) and 5(c) show the calculated reflections from left side $R_L$ and right side $R_R$ in the log scale, respectively. For comparsion, we plot two boundaries of EHS in $(p,q)$ space from Fig. 1(e) by black dashed rings. It can be seen that $R_L$ is enhanced on the outer ring and reduced on the inner ring. On the contrary, $R_R$ is reduced on the outer ring and enhanced on the inner ring. Correspondingly, the unitary transmission is obtained from both sides which is protected by reciprocal symmetry, and this phenomenon is referred as anisotropic transmission resonance (ATR) [51]. With increase of $\gamma$, the inner ring will shrink and the outer ring will expand. At a particular $\gamma$, the outer ring touches the boundary and the inner ring shrinks to one point, which corresponds to the EDPs defined in Fig. 3(b).

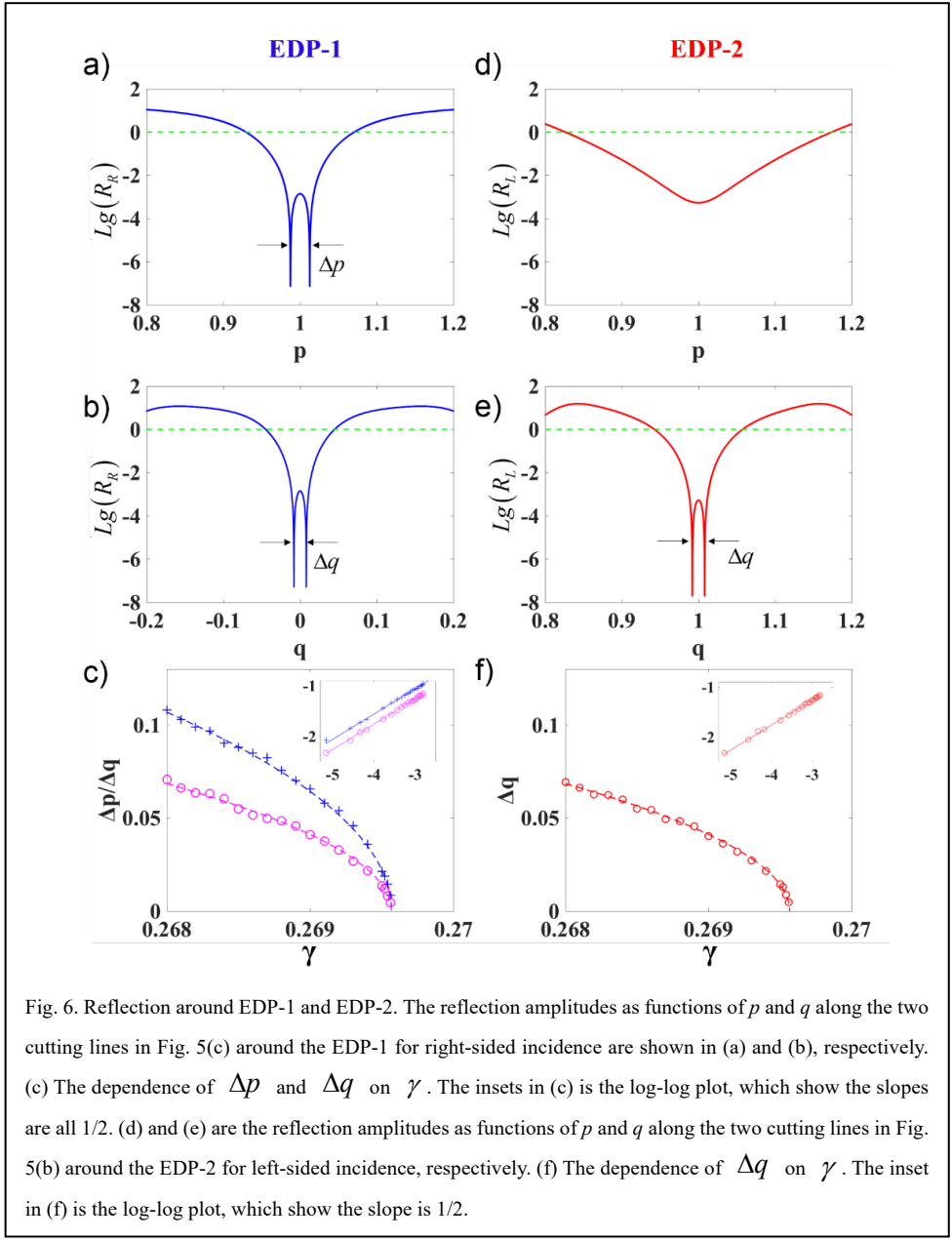

Fig. 6. Reflection around EDP-1 and EDP-2. The reflection amplitudes as functions of $p$ and $q$ along the two cutting lines in Fig. 5(c) around the EDP-1 for right-sided incidence are shown in (a) and (b), respectively. (c) The dependence of $\Delta p$ and $\Delta q$ on $\gamma$. The insets in (c) is the log-log plot, which show the slopes are all 1/2. (d) and (e) are the reflection amplitudes as functions of $p$ and $q$ along the two cutting lines in Fig. 5(b) around the EDP-2 for left-sided incidence, respectively. (f) The dependence of $\Delta q$ on $\gamma$. The inset in (f) is the log-log plot, which show the slope is 1/2.

The EDPs found in the scattering calculation here happen at around $\gamma_0 = 0.2698$, but the EDPs obtained from band structure calculations is at $\gamma = 0.249$. The small deviation is due to the finite size effect.

To further confirm EDPs, we plot the calculated $R_R$ on the $q=0$ axis and the $p=1$ axis near EDP-1, as shown in Figs. 6(a) and 6(b). We see two reflection dips near the EDP-1 in both axes because EDP-1 is a vertical exceptional cone (Fig. 3(d)). We label the distance between these two dips as $\Delta p$ and $\Delta q$. Near the EDP-1, two dips are close to each other and this makes a narrow plateau between them. When $\gamma$ approaches the critical value $\gamma_0$, $\Delta p$ and $\Delta q$ will approach zero, as shown by blue dots and red circles in Fig. 4(c). The inset in Fig. 6(c) gives the dependence of $\log(|\gamma-\gamma_0|)$ on $\log(\Delta p)$ and $\log(\Delta q)$, which is linearly fitted with slope 1/2. This slope further confirms the Hamiltonian in Eq. (3). In Fig. 7(a), we also plot the reflection of the PCs with q=0. When changing the value of $\gamma$ (marked by different color), we can clearly see the two dips fuse and then disappear. The two dips coalesce between $\gamma \in (0.2695, 0.2697)$.

Similar phenomena are observed near EDP-2 as shown in Figs. 6(d) and 6(e) for the change of reflection $R_L$ on the $q=1$ axis and $p=1$ axis. Since EDP-2 is a horizontal Dirac cone, so only the axis $p=1$ intersect with the dashed line of ATR and the axis $q=1$ does not intersect with dashed line before $\gamma$ reaching $\gamma_0 = 0.2695$. Therefore, we find two reflection dips on the p=1 axis, and no reflections dip on the q=1 axis. When $\gamma$ approach the critical value $\gamma_0$, the value of $\Delta q$ will approach zero as shown in Fig. 6(f). The inset in Fig. 6(f) gives the dependence of $\log(|\gamma-\gamma_0|)$ on $\log(\Delta q)$, which is linearly fitted with slope 1/2. This also confirms the Hamiltonian in Eq. (3) of main text.

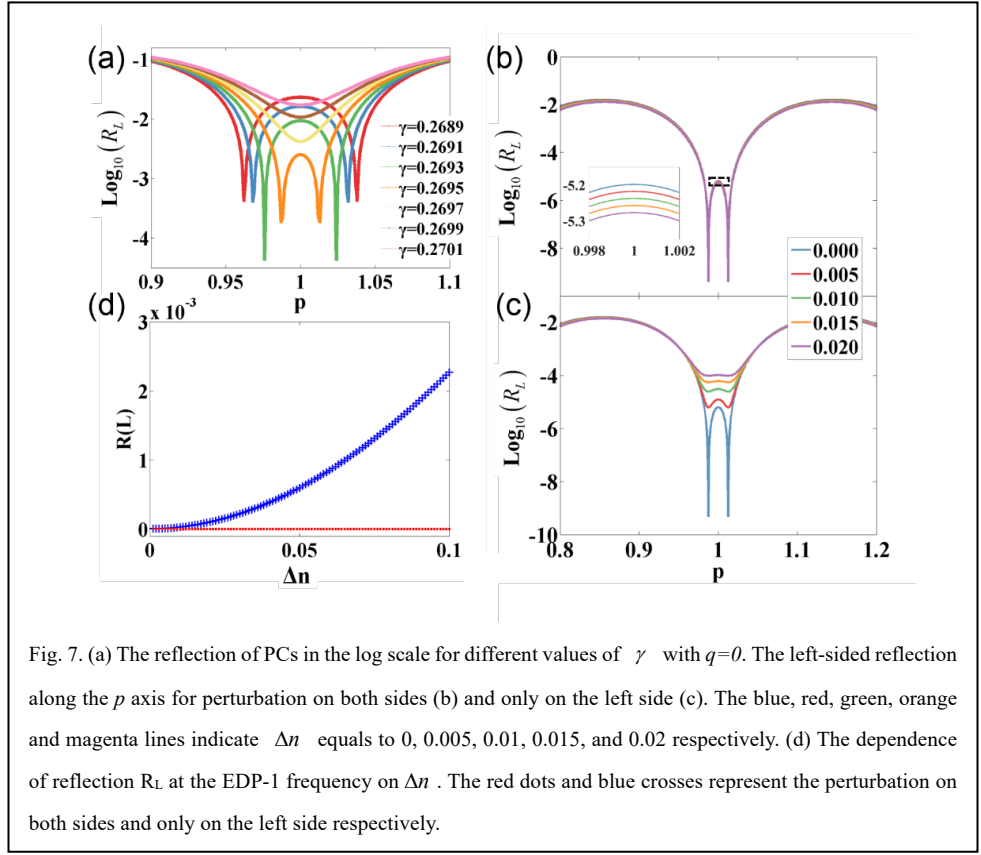

Fig. 7. (a) The reflection of PCs in the log scale for different values of $\gamma$ with $q=0$. The left-sided reflection along the $p$ axis for perturbation on both sides (b) and only on the left side (c). The blue, red, green, orange and magenta lines indicate $\Delta n$ equals to 0, 0.005, 0.01, 0.015, and 0.02 respectively. (d) The dependence of reflection $R_L$ at the EDP-1 frequency on $\Delta n$. The red dots and blue crosses represent the perturbation on both sides and only on the left side respectively.

Since small difference between environment index on two sides will break the PT-symmetry, so the EHS here can be used as optical sensing [23-25, 43]. We consider the change of reflection plateau induced by the perturbation of environment index of PC near the EDP-1. For the system in Fig. 5(a), we firstly consider the environment index on both sides changed with the same value $n_L = n_R = 1 + \Delta n$. Figure 7(b) shows reflections on the axis $q=0$ in the log scale for different index change $\Delta n$. From the inset of Fig. 7(b), we see the change is very small. This shows that the reflection in this senario is not sensitive to the enviroment index. However, if we change refractive index as $n_L = 1 + \Delta n$ and $n_R = 1$, then the reflections are shown in Fig. 7(c). Compared with the results in Fig. 4(d), the change of reflection plateau under different $\Delta n$ is distinguished. The dependence of reflection at EDP-1 for these two cases are given in Fig. 7(d) with red dots (with PT symmetry) and blue crosses (without PT symmetry). This shows that the reflection is quite sensitive to the environment perturbation which breaks PT-symmetry of the system.

## 6. Discussion

Previously, the synthetic dimension must be a degree of freedom, which can be used to simulate the behavior of a spatial dimension, for example, the 4D quantum hall effect [2, 52-54]. However, the concept of the synthetic dimension has been extended very recently to the scenario that an independent systematic parameter can also be regarded as a 'synthetic dimension'[1, 10], which is not necessary to mimic a real spatial dimension. We can introduce the synthetic dimension by using the parameter to either construct a lattice or tune the system. Similarly, we have introduced two geometric parameter *p*, *q* and the non-Hermitian term $\gamma$ in our work. Although the $\gamma$ cannot behave like a spatial dimension, it actually is a new dimension (independent of the other three dimensions). And we can exploit the physics in the Non-Hermitian 4D parameter space.

In Hermitian system, when the topological phase transition occurs, the bands will always have degenerate points, for example, the Dirac cone behaves as the phase transition point. However, the degenerate point of exceptional surface has not been discussed before. Here we have observed the exceptional cone in the 4D parameter space firstly, which behaves topological phase transition of the exceptional surface. The exceptional cone is a linearly degenerate point of the exceptional surface. This is quite different from the previous case, that the linear crossing point of exceptional lines is always not an EP, because the two EPs with opposite chirality coalesce at this point, as shown in Fig. 1(f)[45, 46].

## 7. Conclusion

Through introducing two parameters (p, q), we construct 4D parameter space $(p, q, k, \gamma)$ in a 1D PC with a complex unit cell. Based on the symmetry of parameter space, we obtain exceptional cone in the EHS. Through calculation of the geometric Phase, we prove such degeneracy points are the consequence of topological properties of EPs. We further show such degeneracy point is quite sensitive to environment perturbation which breaks PT-symmetry and can be possibly used in optical sensing. Our work shows that parameter space is a very powerful method to explore the high dimensional non-Hermitian topological systems. With the help of non-Hermitian parameters, the 1D PC with complex unit cell is a very flexible experimental platform to mimic topological properties of non-Hermitian exceptional hypersurfaces. Some other optical properties of this kind of degeneracy point are worthy to be investigate in the future. Beside optical sensing proposed in this work, this degeneracy point can be possibly applied in other nonlinear and quantum optical processes.

**Funding.** National Natural Science Foundation of China (Nos. 11690033, 61425018, 11621091), National Key Research and Development Program of China (No.

2017YFA0205700), National Key R&D Program of China (2017YFA0303702), Research Grants Council of Hong Kong (grant no. AoE/P-02/12 and N_HKUST608/17) and William Mong Institute of Nano Science and Technology，Gordon and Betty Moore Foundation，Singapore Ministry of Education (grant MOE2015-T2-2-008).

**Disclosures.** The authors declare no conflicts of interest.